\documentclass[%
 aip,
 pof,
 floatfix,
 amsmath,amssymb,
 reprint,%
]{revtex4-1}
\usepackage{graphicx}% Include figure files
\usepackage{dcolumn}% Align table columns on decimal point
\usepackage{bm}% bold math
\usepackage[utf8]{inputenc}
\usepackage[T1]{fontenc}
\usepackage{mathptmx}
\usepackage{etoolbox}
\usepackage{booktabs}
\usepackage{longtable}
\usepackage{tabularx}

\makeatletter
\def\@email#1#2{%
 \endgroup
 \patchcmd{\titleblock@produce}
  {\frontmatter@RRAPformat}  {\frontmatter@RRAPformat{\produce@RRAP{*#1\href{mailto:#2}{#2}}}\frontmatter@RRAPformat}
  {}{}
}%
\makeatother

\begin{document}
\preprint{AIP/123-QED}
\title[Physics of Fluids]{Spontaneous translation of charged droplets during evaporation on dry surfaces}

\author{Riming Xu}

\affiliation{Physical Science and Engineering, King Abdullah University of Science and Technology (KAUST), Thuwal 23955, Saudi Arabia}

\author{Yanbo Li}
\affiliation{Department of Civil and System Engineering, Johns Hopkins University, Baltimore 21218,
USA.}

\author{Jiawen Zhang}
\affiliation{Department of Materials Science and Engineering, Stanford University, Stanford, CA, 94305,
USA}

\author{Jin Wang}
\thanks{Jin Wang and Yikai Li are the corresponding author for this work.}
\affiliation{Physical Science and Engineering, King Abdullah University of Science and Technology (KAUST), Thuwal 23955, Saudi Arabia}

\author{Yikai Li}
\thanks{Jin Wang and Yikai Li are the corresponding author for this work.}
\email{<wangjinbjut321@163.com>, <liyikai@bit.edu.cn>}
\affiliation{School of Mechanical Engineering, Beijing Institute of Technology, Beijing, China}

\date{\today}% It is always \today, today,

\begin{abstract}
Evaporating sessile droplets are usually treated as capillary objects, but droplets generated by routine handling can carry tens to hundreds of picocoulombs of electric charge. Here we combine Faraday-cup charge measurements with optical imaging to determine how such charge evolves as water droplets evaporate on dry polymer substrates. A zero-time protocol shows that a reproducible initial charge is preserved on poly(methylpentene) (PMP), whereas PDMS, SOCAL-coated surfaces, and polystyrene either exchange, dissipate, or inject charge on contact. On PMP, ensemble-resolved measurements reveal two regimes: the charge remains nearly constant during early evaporation and then decreases abruptly once the droplet reaches a small-volume state. This charge collapse coincides with spontaneous lateral translation rather than jetting or breakup. A Rayleigh-normalized analysis, including a spherical-cap stress correction and measured contact-angle retention scale, shows that motion occurs only after evaporation drives the droplet into a high electro-pinning state. High-speed imaging and kinematic analysis support a picture in which the subsequent motion is governed by repeated contact-line depinning and re-pinning: the total distance traveled is strongly affected by dry-surface pinning, whereas the peak translational velocity serves as a more robust indicator of the discharge strength. These results identify a dry-substrate mode of evaporation-driven electrostatic relaxation, distinct from Coulomb fission on lubricated surfaces, in which substrate electrostatic passivity enables charge retention, droplet geometry selects the instability onset, and whole-droplet translation provides the charge-release pathway.
\end{abstract}

\maketitle

\section{\label{sec:level1}Introduction}

Evaporating sessile droplets are a central model system for non-equilibrium interfacial transport. Their lifetime, internal flow, contact-line motion, and deposition patterns are controlled by evaporation, capillarity, wetting, and substrate interactions, which has made them important in coating, printing, heat transfer, microfluidics, aerosol science, and pattern formation \cite{picknett1977evaporation,deegan1997capillary,deegan2000contact,hu2002evaporation,popov2005evaporative,stauber2015lifetimes,wilson2023evaporation}. In this classical view, the droplet is treated mainly as a capillary object: evaporation changes the geometry, capillary stresses drive flow, and contact-angle hysteresis regulates pinning and depinning. Electrical charge is usually treated as a secondary effect, or is omitted altogether.

This simplification is not always justified. Charged water droplets are common in natural and technological settings, from cloud droplets and charged evaporation aerosols to sprays, electrospray ionization, inkjet processes, and laboratory liquid handling \cite{takahashi1973measurement,tinsley2000effects,tan2016enhanced,iribarne1976evaporation,kebarle2000mechanisms}. Even routine pipetting can generate droplets carrying tens to hundreds of picocoulombs, and liquid--solid contact can transfer charge between water and insulating surfaces \cite{choi2013spontaneous,sun2016solid,baytekin2011mosaic,lacks2011contact,lin2022contactreview}. Once present, this charge is not merely a passive label: it can alter droplet impact, jumping, sliding, coalescence, and surface transport, and it can couple to interfacial charge distributions created by wetting or contact electrification \cite{miljkovic2013electrostatic,nauruzbayeva2020electrification,li2022charge,sun2019surface,xu2021impact,diaz2023self}. These observations suggest that evaporation on insulating substrates may involve a coupled electro-capillary state rather than a purely hydrodynamic one.

Evaporation is especially important for charged droplets because it changes the ratio of electrostatic stress to capillary stress. For an isolated conducting droplet, Rayleigh showed that electrostatic repulsion destabilizes the interface when the charge exceeds a size-dependent limit \cite{rayleigh1882liquid}. This idea underlies extensive work on electrosprays and ion formation, where evaporating charged droplets may undergo repeated Rayleigh fission, ion evaporation, or charge-residue pathways \cite{iribarne1976evaporation,fenn1993ion,kebarle2000mechanisms,gomez1994charge,duft2003coulomb}. Single-droplet experiments have confirmed that evaporating charged microdroplets can lose charge abruptly near the Rayleigh limit, while spray and breakup studies reveal that charge release can occur through jets, progeny droplets, or charge-separated fragments \cite{li2005charge,zilch2008charge}. More recently, charged water drops on lubricated, low-pinning surfaces were shown to undergo spontaneous Coulomb fissions during evaporation, with periodic elongation, jetting, and microdroplet ejection \cite{lin2026spontaneous}. Together, these studies establish evaporation-driven charge concentration as a route to interfacial instability.

The corresponding problem for dry sessile droplets remains less clear. A dry solid substrate introduces features absent from free, suspended, sprayed, or lubricated droplets: direct liquid--solid charge exchange, heterogeneous surface charge, contact-line pinning, defect-mediated depinning, and possible charge leakage through the substrate. Measurements of charged droplets in dielectric liquids and on electrode surfaces show that droplet charge can depend sensitively on contact, field strength, droplet size, electrolyte content, and the measurement pathway itself \cite{jung2008electrical,im2011electrophoresis,im2012discrete}. Thus, before asking how evaporation concentrates charge, one must first ask whether a dry substrate preserves a well-defined initial charge state at all. Without such a state, charge relaxation cannot be separated from instantaneous charging artifacts, substrate-mediated charge injection, or rapid dissipation.

Here we investigate evaporation-driven charge dynamics of water droplets on dry polymer substrates with distinct electrostatic behavior: poly(methylpentene) (PMP), polydimethylsiloxane (PDMS), SOCAL-coated surfaces, and polystyrene. We use a custom Faraday-cup protocol with matched zero-time controls to separate the charge introduced before deposition from charge exchanged during contact, retrieval, and evaporation. This protocol allows us to address three connected questions. First, which dry substrates preserve a reproducible droplet charge after deposition? Second, when a charge-retentive substrate exists, does evaporation produce gradual charge leakage or a geometry-selected instability? Third, if the charged droplet becomes unstable on a dry pinned surface, does it relax by the Coulomb-fission pathway known from lubricated surfaces, or by a different dry-surface mode?

We find that PMP is unique among the tested substrates in preserving a reproducible initial charge while also supporting a late-stage electrostatic instability. On PMP, the droplet charge remains nearly constant through early evaporation and then collapses sharply only after the droplet reaches a small-volume, highly pinned state. This charge loss coincides with spontaneous lateral translation and repeated contact-line depinning and re-pinning, rather than with jetting or breakup. By contrast, PDMS, SOCAL, and polystyrene do not show the same coupled behavior because they respectively introduce heterogeneous charge exchange, rapid dissipation, or additional substrate-to-droplet charging.

These results define a dry-substrate pathway for evaporation-driven electrostatic relaxation. The onset is selected by droplet geometry and contact-line retention: evaporation increases the electro-capillary stress until it exceeds the measured pinning scale. After onset, the total travel distance is strongly affected by dry-surface defects and re-pinning, whereas the peak translational velocity correlates more directly with the charge relaxed during the event. Spontaneous translation on PMP is therefore distinct from lubricated-surface Coulomb fission: the charged droplet remains intact, but releases electrostatic stress through whole-droplet motion on a dry insulating substrate.

\section{Experimental system and methods}
\subsection{Experimental setup}

Experiments were performed using sessile water droplets deposited on solid polymer substrates under ambient laboratory conditions. Droplets with an initial volume of $0.6~\mu\mathrm{L}$ were gently placed onto the substrate using standard liquid-handling tools, including plastic pipettes, glass tubes, or metal needles, depending on the specific experiment.

The substrates investigated in this study included untreated poly(methylpentene) (PMP), polydimethylsiloxane (PDMS), SOCAL-coated surfaces, and polystyrene. PMP substrates were commercially sourced (Thermo Scientific, Nalgene series) and used as received. All substrates were employed in their dry state without additional lubrication or surface treatment. Prior to each experiment, substrates were cleaned with ethanol and deionized water and allowed to dry in air. Each substrate was mounted on an electrically isolated holder to minimize unintended charge leakage.

The net electric charge of individual droplets was measured using a custom-built Faraday cup connected to a high-sensitivity electrometer (Keithley 6514). At a prescribed time $t$, the droplet was carefully retrieved from the substrate and transferred into the Faraday cup, where the total charge was recorded. The electrometer provides a charge resolution on the order of $10^{-14}\,\mathrm{C}$, and the overall measurement uncertainty of the Faraday cup system was estimated to be approximately $\pm 0.01\,\mathrm{pC}$. This uncertainty is several orders of magnitude smaller than the typical droplet charges measured in this study, which range from tens to hundreds of picocoulombs, ensuring that instrumental noise does not limit the reported results. Because the Faraday-cup measurement irreversibly discharges the droplet, $Q_i$ and $Q_t$ were not measured sequentially on the same droplet. Instead, the normalized charge $Q_t/Q_i$ was reconstructed from matched independent trials: zero-time reference measurements defined the initial charge distribution $Q_i$, while separate droplets prepared under the same conditions were measured after a prescribed residence time to obtain $Q_t$.

\begin{figure}[!t]
  \centering
  \includegraphics[width=0.5\linewidth]{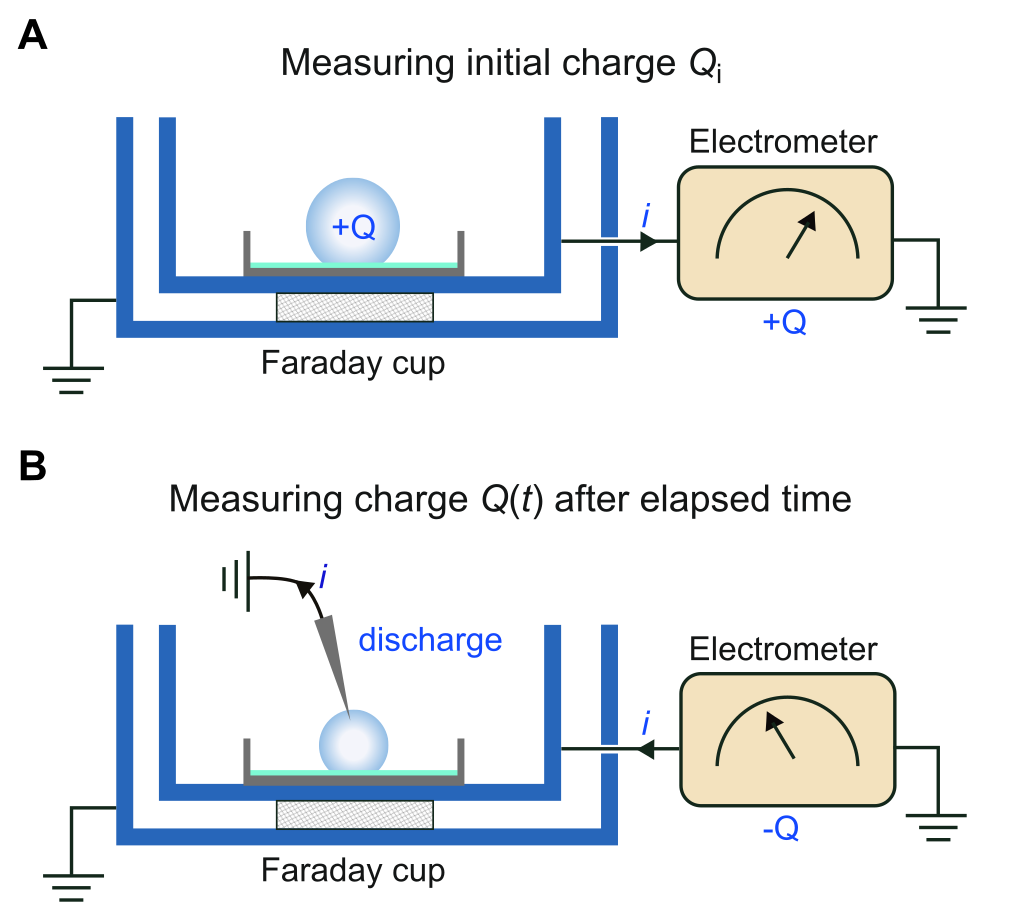}
\caption{Schematic of the charge measurement procedure using a Faraday cup.
(A) Zero-time reference measurement of the initial charge, $Q_i$. A freshly prepared charged droplet is transferred into the Faraday cup, and the net charge is recorded by an electrometer.
(B) Charge measurement after an elapsed evaporation time, $Q_t$. In a matched independent trial, a freshly prepared droplet resides on the substrate for a prescribed duration before retrieval into the Faraday cup. Because each measurement irreversibly discharges the droplet, the reported $Q_t/Q_i$ values are ensemble-matched snapshots rather than repeated measurements of a single droplet.}
  \label{fig:setup}
\end{figure}

For selected experiments, side-view optical imaging was employed to monitor droplet shape and dynamics during evaporation. Images were acquired using a high-resolution camera equipped with a long-working-distance objective and were used to extract the droplet width-to-height ratio (W/H), apparent contact angle, and stationary/moving state. The apparent contact angle was obtained from spherical-cap fits to the side-view profiles. In a separate set of PMP side-view measurements immediately before spontaneous motion, the apparent angle decreased from $104.2^\circ$ at deposition to $74.7^\circ$ before motion, corresponding to a representative pinning term $\Delta\cos\theta_{\mathrm{pin}}\approx0.51$. Top-view high-speed recordings were analyzed to obtain the maximum travel distance $D_{\max}$ and maximum translational speed $U_{\max}$ for individual translation events. A schematic illustration of the experimental setup and the charge measurement procedure is shown in Fig.~\ref{fig:setup}.

\subsection{Charge measurement protocol and validation}

The net electric charge of individual droplets was quantified using the Faraday cup protocol illustrated in Fig.~\ref{fig:setup}. For each measurement, the droplet was physically transferred into the Faraday cup, where the induced signal was recorded by the electrometer. This process irreversibly discharges the droplet, precluding repeated charge measurements on the same sample. Consequently, all charge measurements reported in this work were obtained from independent experimental trials, and temporal charge evolution should be interpreted as an ensemble trajectory reconstructed from matched snapshots.

To assess the robustness of the charging and measurement procedure, droplets were generated using three different liquid-handling methods: plastic pipettes, glass tubes, and metal needles. These methods span a wide range of contact electrification conditions. For all three cases, the measured initial charge distributions were comparable, and no systematic dependence of the subsequent charge evolution on the charging method was observed. This confirms that the experimental observations reported here do not depend sensitively on the specific charging mechanism.

To further exclude measurement artifacts and instantaneous charge loss upon contact with the substrate, a zero-time control experiment was performed. In this protocol, a charged droplet was deposited onto the substrate and immediately retrieved and transferred into the Faraday cup, eliminating evaporation and long-time contact. On PMP, the measured zero-time charge after retrieval matched the independently measured initial charge within experimental uncertainty over a broad range of charge magnitudes. This result demonstrates that neither the Faraday cup measurement nor instantaneous contact with PMP induces systematic charge dissipation, which justifies using PMP as the reference substrate for the long-time evaporation measurements.

Taken together, these validation experiments confirm that the observed charge decay during evaporation arises from time-dependent droplet--substrate interactions rather than from instrumental limitations or measurement-induced artifacts.

\section{Substrate screening and zero-time charge controls}

\subsection{PMP as a charge-retentive reference substrate}

\begin{figure}[!t]
  \centering
  \includegraphics[width=0.8\linewidth]{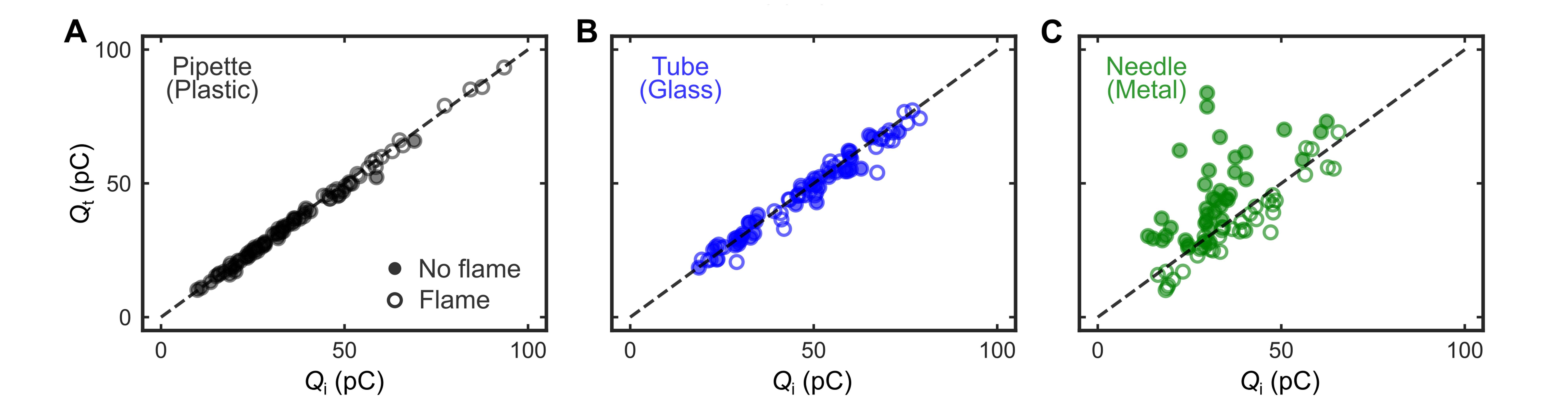}
  \caption{Comparison of zero-time charge measurements on PMP using different measurement approaches.
  (A) Charge extraction using a plastic pipette.
  (B) Charge extraction using a glass tube.
  (C) Charge measurement via contact with a grounded fine metal needle without removing the droplet.
  In all panels, the measured charge after immediate interaction with the substrate, $Q_t$, is plotted against the initial charge, $Q_i$.
  The dashed line indicates $Q_t = Q_i$.
  Filled symbols correspond to experiments performed on untreated PMP substrates, whereas open symbols represent measurements conducted after flame treatment of the PMP surface to reduce residual static charge.
  }
  \label{fig:zero_time}
\end{figure}

Before analyzing evaporation-driven dynamics, it is necessary to determine whether a substrate preserves a meaningful initial droplet charge at all. Figure~\ref{fig:zero_time} shows zero-time controls on PMP, in which droplets were deposited and immediately retrieved before measurable evaporation occurred. Across plastic-pipette extraction, glass-tube extraction, and grounded-needle discharge, the measured charge is compared with the independently measured initial charge. The purpose of this test is not to study long-time evaporation, but to identify whether PMP can serve as a clean platform for subsequent charge-retention measurements.

When a plastic pipette is used to extract the droplet (Fig.~\ref{fig:zero_time}(a)), the measured charges collapse tightly around the $Q_t = Q_i$ line with minimal scatter. This result demonstrates that plastic pipette tips do not induce measurable charge transfer during extraction. Despite the inevitable motion of the three-phase contact line during droplet removal, no systematic deviation from charge conservation is observed. This indicates that contact-line motion alone does not lead to additional charge acquisition by the droplet, and that there is no detectable charge exchange between the droplet and the PMP substrate in the zero-time limit.

The glass-tube and grounded-needle controls clarify the main experimental artifacts. Glass extraction preserves the mean charge but introduces larger scatter, consistent with static charge on the tube. Needle discharge on untreated PMP is more sensitive to residual substrate charge, whereas flame-treated PMP collapses back toward the $Q_t=Q_i$ reference. Taken together, these controls establish PMP as the only tested dry substrate on which the initial charge state is reproducible enough to support a geometry-resolved evaporation study.

\subsection{Electrostatically active control substrates}

The same zero-time protocol was then applied to PDMS, SOCAL surfaces, and polystyrene. These control measurements show why the long-time instability analysis should focus on PMP rather than on all substrates equally.

\begin{figure}[!t]
  \centering
  \includegraphics[width=0.8\linewidth]{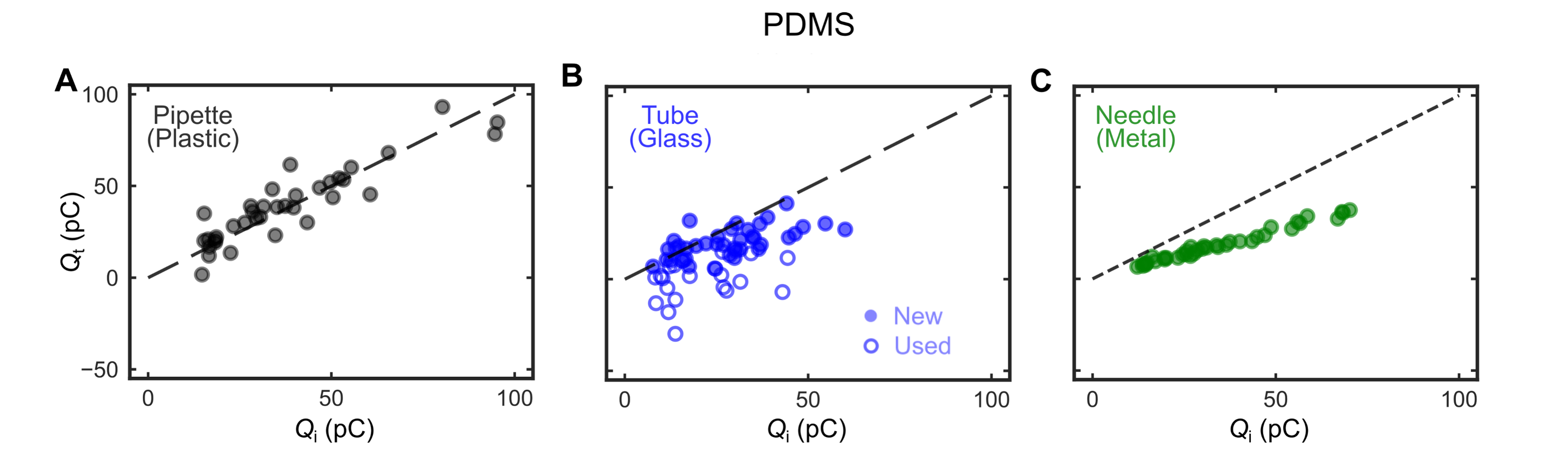}
  \caption{Zero-time charge measurements on PDMS substrates using different measurement approaches.
  (A) Charge extraction using a plastic pipette.
  (B) Charge extraction using a glass tube.
  (C) Charge measurement via contact with a grounded metal needle without removing the droplet.
  In all panels, the measured charge $Q_t$ is plotted against the initial charge $Q_i$, with the dashed line indicating $Q_t = Q_i$.
  Filled symbols denote measurements performed using newly prepared tools, whereas open symbols correspond to measurements conducted using reused glass tubes.
  }
  \label{fig:pdms_zero_time}
\end{figure}

On PDMS, the recovered charge exhibits large trial-to-trial scatter even in the zero-time limit (Fig.~\ref{fig:pdms_zero_time}). This behavior is consistent with the mosaic-like surface charge distribution known to arise during contact electrification of polymer surfaces \cite{baytekin2011mosaic}. Reused glass tubes further amplify the scatter by carrying their own static charge, and needle discharge is affected by substrate-bound charge. Thus, PDMS does not provide a uniquely defined initial droplet charge state.

\begin{figure}[!t]
  \centering
  \includegraphics[width=0.8\linewidth]{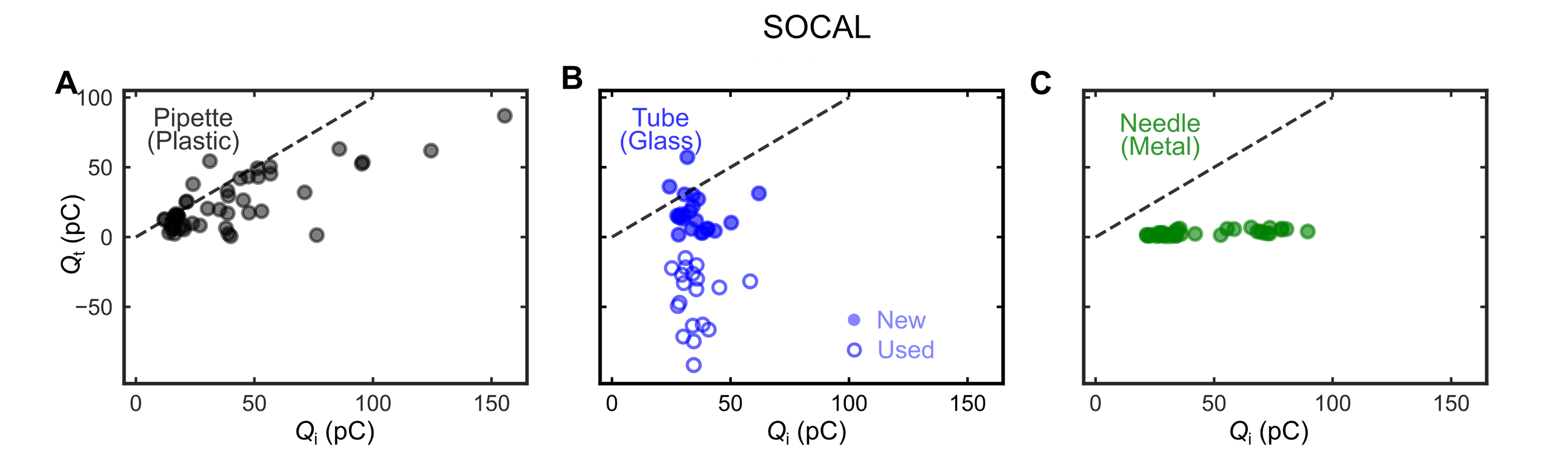}
  \caption{Zero-time charge measurements on SOCAL-coated substrates using different measurement approaches.
  (A) Charge extraction using a plastic pipette.
  (B) Charge extraction using a glass tube.
  (C) Charge measurement via contact with a grounded metal needle without removing the droplet.
  In all panels, the measured charge $Q_t$ is plotted against the initial charge $Q_i$, with the dashed line indicating $Q_t = Q_i$.
  The results demonstrate the combined influence of contact line motion, surface-mediated charge generation, and rapid charge dissipation on SOCAL substrates.
  }
  \label{fig:socal_zero_time}
\end{figure}

SOCAL represents a different failure mode (Fig.~\ref{fig:socal_zero_time}). Pipette and glass-tube extraction indicate charge generation during contact-line motion on the surface, while grounded-needle measurements collapse toward zero charge, demonstrating rapid charge dissipation. SOCAL therefore acts both as an active liquid--solid charging interface and as an efficient charge sink, preventing the charge-retentive pathway observed on PMP.

\begin{figure}[!t]
  \centering
  \includegraphics[width=0.8\linewidth]{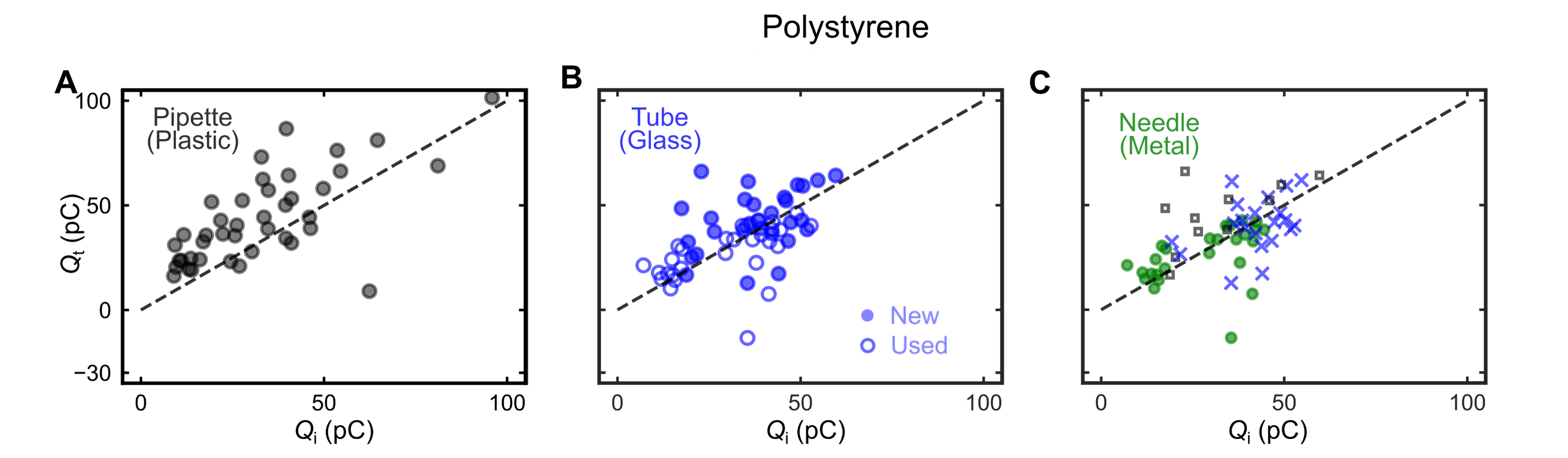}
  \caption{Zero-time charge measurements on polystyrene substrates using different measurement approaches.
  (A) Charge extraction using a plastic pipette.
  (B) Charge extraction using a glass tube.
  (C) Charge measurement via contact with a grounded metal needle without removing the droplet.
  In all panels, the measured charge $Q_t$ is plotted against the initial charge $Q_i$, with the dashed line indicating $Q_t = Q_i$.
  The data exhibit substantial trial-to-trial variability, indicating that polystyrene does not support a uniquely defined initial droplet charge state and that charge exchange between the droplet and the substrate depends sensitively on local surface conditions.
  }
  \label{fig:polystyrene_zero_time}
\end{figure}

Polystyrene also fails to preserve a clean initial state (Fig.~\ref{fig:polystyrene_zero_time}). The recovered charge varies strongly between trials and can deviate substantially from the $Q_t=Q_i$ reference, indicating local charge exchange between the droplet and the substrate. These three control substrates therefore show that evaporation-induced shape change alone is not sufficient to produce a useful charge-instability experiment. A substrate must first be electrostatically passive enough to retain droplet charge; among the materials tested here, PMP uniquely satisfies this requirement.

\begin{figure*}[!t]
  \centering
  \includegraphics[width=0.9\linewidth]{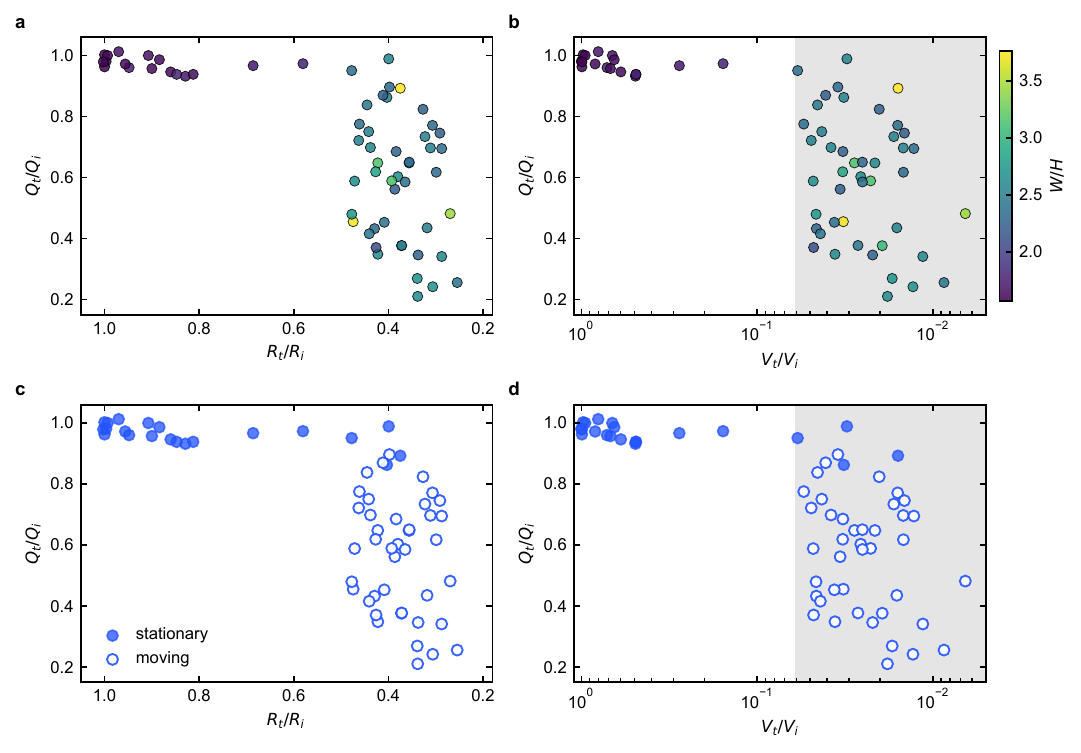}
  \caption{Charge evolution and droplet dynamics during evaporation on PMP substrates.
  (a) Normalized charge $Q_t/Q_i$ as a function of the normalized droplet radius $R_t/R_i$, with the color indicating the instantaneous shape ratio $W/H$.
  (b) The same charge data plotted against the normalized droplet volume $V_t/V_i$ on a logarithmic horizontal axis.
  (c,d) Corresponding state-resolved plots in which filled markers denote stationary droplets and open markers denote droplets undergoing spontaneous lateral motion.
  The shaded regions in (b,d) mark the late small-volume regime in which charge loss and droplet motion emerge together.
  }
  \label{fig:charge_evap_pmp}
\end{figure*}

\section{Evaporation-driven charge instability on PMP}

\subsection{Charge evolution during evaporation on PMP substrates}

Figure~\ref{fig:charge_evap_pmp} summarizes the coupled evolution of droplet charge, geometry, and dynamics during evaporation on PMP substrates. Because each charge measurement irreversibly discharges the droplet, the charge evolution is reconstructed from a series of independent experiments performed at different extraction times. In each experiment, a fresh droplet with a fixed initial volume of $0.6~\mu$L was deposited on the PMP surface, and the remaining charge $Q_t$ was recorded after an elapsed time $t$ by retrieving the droplet into the Faraday cup. The reference charge $Q_i$ was obtained from matched zero-time trials performed with the same droplet preparation protocol, so $Q_t/Q_i$ represents an ensemble-normalized charge-retention ratio.

Figures~\ref{fig:charge_evap_pmp}(a) and (c) show the normalized charge $Q_t/Q_i$ as a function of the normalized droplet radius $R_t/R_i$. Each data point corresponds to a distinct extraction time, with earlier times yielding larger values of $R_t/R_i$. The color scale in Fig.~\ref{fig:charge_evap_pmp}(a) represents the instantaneous droplet shape quantified by the width-to-height ratio $W/H$. At early stages of evaporation ($R_t/R_i \gtrsim 0.5$), the droplet retains nearly all of its initial charge, and $Q_t/Q_i$ remains close to unity. As evaporation proceeds and the droplet radius decreases below this threshold, a pronounced reduction in $Q_t/Q_i$ is observed. Notably, Fig.~\ref{fig:charge_evap_pmp}(c) reveals that this charge loss coincides with the onset of spontaneous droplet motion, indicating a strong coupling between charge relaxation and droplet mobility.

To further clarify the role of droplet geometry, the same data are replotted in terms of the normalized droplet volume $V_t/V_i$ in Figs.~\ref{fig:charge_evap_pmp}(b) and (d). The droplet volume is estimated by approximating the droplet as a spherical cap,
\[
V = \frac{\pi H}{6}\left(3a^2 + H^2\right),
\]
where $H$ is the instantaneous droplet height and $a$ is the base radius obtained from one half of the measured side-view width. When expressed as a function of $V_t/V_i$, the charge evolution exhibits a much sharper transition: significant charge dissipation is confined to a narrow range of small remaining volumes. This representation is particularly appropriate because, once droplet motion sets in, the base radius $R_t$ undergoes periodic variations associated with repeated depinning and re-pinning of the contact line (see Side View Supplementary Videos~2--6 and Top View Supplementary Videos~7--12). As a result, $R_t/R_i$ alone does not provide a strictly monotonic measure of the evaporation state, whereas the droplet volume $V_t$ decreases monotonically with time and therefore offers a more robust descriptor of the droplet's geometric evolution. As shown in Fig.~\ref{fig:charge_evap_pmp}(d), the sharp transition in $Q_t/Q_i$ expressed in terms of $V_t/V_i$ aligns closely with the emergence of droplet motion, demonstrating that charge loss on PMP is not a continuous consequence of evaporation alone but is instead triggered by a geometry-dependent instability that enables droplet displacement and facilitates charge release.

This geometry-selected transition can be interpreted in terms of a competition between electro-capillary stress and contact-line retention. For an isolated spherical droplet of radius $R$, the Rayleigh charge is
\[
Q_R = 8\pi \left(\varepsilon_0 \gamma R^3\right)^{1/2},
\]
and the corresponding fissility is
\[
X=\frac{Q^2}{64\pi^2\varepsilon_0\gamma R^3}
=\left(\frac{Q}{Q_R}\right)^2,
\]
where $\varepsilon_0$ is the permittivity of free space and $\gamma$ is the liquid--air surface tension \cite{rayleigh1882liquid}. We first evaluate this expression using the spherical-equivalent radius $R_{\mathrm{eq}}=(3V_t/4\pi)^{1/3}$, giving an equivalent-sphere reference $X_{\mathrm{eq}}$. This quantity is useful for comparison with classical Rayleigh scaling, but it should not be interpreted as an exact instability threshold for a pinned sessile droplet. To avoid treating the sessile cap as a free sphere, we also compute a measured-shape electro-capillary stress parameter,
\[
X_{\mathrm{cap}} =
\frac{Q_i^2 R_c}{4\varepsilon_0 \gamma A_{lv}^2},
\]
where $A_{lv}=\pi(a^2+H^2)$ is the liquid--air area of a spherical cap with base radius $a$ and height $H$, and $R_c=(a^2+H^2)/(2H)$ is its radius of curvature. This expression compares the electrostatic pressure scale $Q_i^2/(2\varepsilon_0 A_{lv}^2)$ with the curvature pressure scale $2\gamma/R_c$. Although it reduces to the classical Rayleigh fissility for a free spherical droplet, here $X_{\mathrm{cap}}$ is used only as a sessile-drop stress parameter. Thus, $X_{\mathrm{cap}}>1$ indicates that the estimated electrostatic pressure scale exceeds the capillary curvature-pressure scale for the measured cap geometry; it does not imply that the dry sessile droplet has crossed the free-drop Rayleigh threshold. In the present data $a$ is one half of the measured side-view width, so the calculation uses the actual liquid--air area and curvature of the evaporating cap rather than only its volume.

The remaining question is whether the electro-capillary stress is large enough to overcome contact-line retention. For a pinned droplet, the lateral retention force is commonly scaled as $\gamma \Delta\cos\theta$ per unit contact-line length, where $\Delta\cos\theta=\cos\theta_R-\cos\theta_A$ represents contact-angle hysteresis \cite{furmidge1962studies,deGennes1985wetting,joanny1984contact,eral2013contact}. Direct side-view contact-angle traces on PMP give an apparent angle of $104.2^\circ$ shortly after deposition and $74.7^\circ$ immediately before spontaneous motion, corresponding to $\Delta\cos\theta_{\mathrm{pin}}\approx0.51$. We therefore define an electro-pinning number
\[
\Pi_{\mathrm{dep}} =
\frac{Q_i^2/(2\varepsilon_0 A_{lv}^2)}
{\gamma\Delta\cos\theta_{\mathrm{pin}}/a}
=
\frac{2a}{R_c}
\frac{X_{\mathrm{cap}}}{\Delta\cos\theta_{\mathrm{pin}}}.
\]
Values of $\Pi_{\mathrm{dep}}$ above unity indicate that the electrostatic stress scale exceeds the measured capillary retention scale. This is not a microscopic model for individual defects, but it provides a quantitative link between fissility, pinning/depinning, and charge relaxation using measured geometry and measured wetting hysteresis.

\begin{figure*}[!t]
  \centering
  \includegraphics[width=0.9\linewidth]{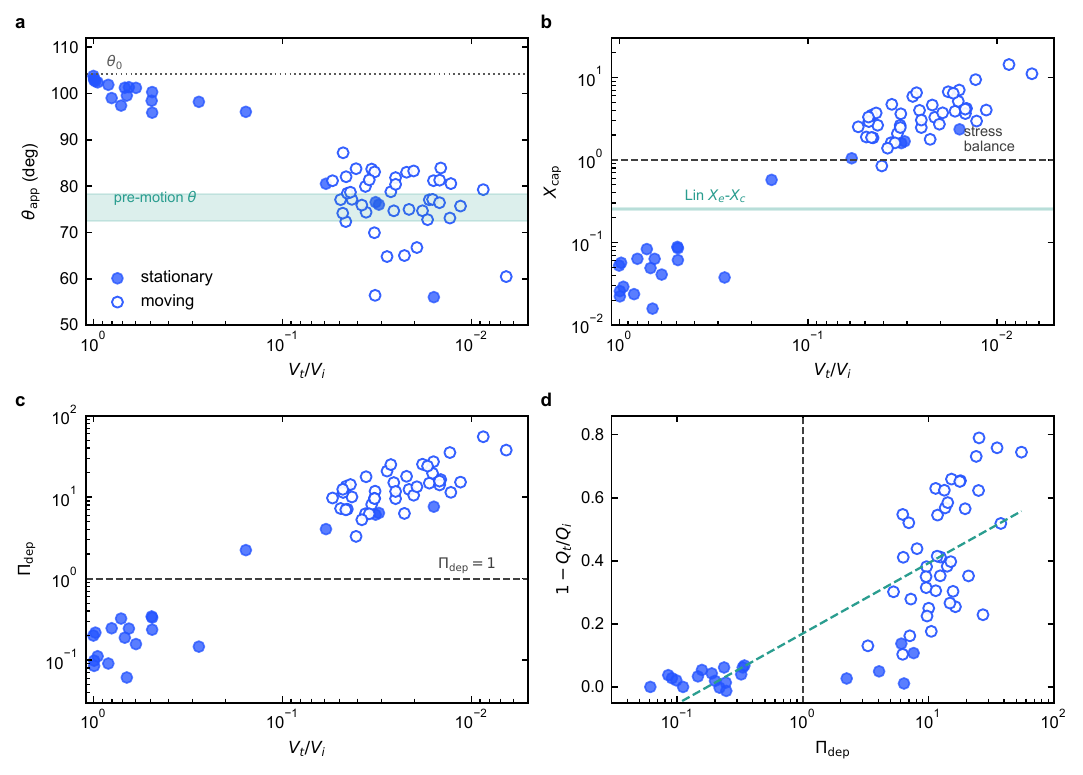}
  \caption{Electro-capillary and pinning-controlled onset of spontaneous translation on PMP.
  (a) Apparent contact angle $\theta_{\mathrm{app}}$ calculated from the measured spherical-cap geometry as a function of normalized volume $V_t/V_i$. The dotted line gives the median direct initial contact angle on PMP, and the shaded band gives the measured pre-motion angle range from independent side-view contact-angle traces.
  (b) Measured-shape electro-capillary stress parameter $X_{\mathrm{cap}}=Q_i^2R_c/(4\varepsilon_0\gamma A_{lv}^2)$. This quantity is a sessile-cap stress ratio, not a free-drop Rayleigh threshold. The cyan band marks the critical fissility range $X_e=0.25$ to $X_c=0.26$ reported by Lin \textit{et al.} for elongation and Coulomb fission of lubricated sessile droplets; their modified fissility uses an effective interfacial tension and hemispherical lubricated-drop geometry, so the band is included as a scale comparison rather than as an identical threshold for $X_{\mathrm{cap}}$.
  (c) Electro-pinning number $\Pi_{\mathrm{dep}}=2(a/R_c)X_{\mathrm{cap}}/\Delta\cos\theta_{\mathrm{pin}}$, where $\Delta\cos\theta_{\mathrm{pin}}=0.51$ is obtained from the contact-angle measurements.
  (d) Relaxed charge fraction $1-Q_t/Q_i$ plotted against $\Pi_{\mathrm{dep}}$.
  Filled markers denote stationary droplets and open markers denote droplets that undergo spontaneous lateral motion. Dashed lines mark electro-capillary stress balance for $X_{\mathrm{cap}}$ and depinning balance for $\Pi_{\mathrm{dep}}$.
  }
  \label{fig:pmp_fissility}
\end{figure*}

The resulting phase map is shown in Fig.~\ref{fig:pmp_fissility}. The initial charge $Q_i$ is used as the pre-relaxation charge because the zero-time and early-evaporation measurements show charge retention on PMP; for moving droplets, $Q_t$ is the post-relaxation charge. The wetting data in Fig.~\ref{fig:pmp_fissility}(a) show that droplets approach the independently measured pre-motion contact-angle band only at small $V_t/V_i$. At the same time, the measured-shape electro-capillary stress parameter separates the moving and stationary populations: moving droplets occupy the high-$X_{\mathrm{cap}}$ regime, whereas stationary droplets are mostly clustered near low stress ratios. Because $X_{\mathrm{cap}}$ is a sessile-cap pressure ratio rather than a free-drop Rayleigh threshold, the central onset metric is the electro-pinning number. Incorporating the measured pinning term gives a clearer dimensional comparison: moving droplets have a median $\Pi_{\mathrm{dep}}=12.5$ and all moving events have $\Pi_{\mathrm{dep}}>3.3$, while stationary droplets have a median $\Pi_{\mathrm{dep}}=0.24$. A few stationary points exceed unity, which is expected for a dry substrate because local defects and surface-charge heterogeneity can delay the actual depinning event. Nevertheless, Fig.~\ref{fig:pmp_fissility}(d) shows that large charge relaxation occurs only after the droplet enters the high-$\Pi_{\mathrm{dep}}$ regime. The onset criterion is therefore not simply ``small droplet'' or ``large charge''; it is the combined condition that evaporation increases electro-capillary stress enough to overcome the measured wetting-retention scale.

\begin{figure}[!t]
  \centering
  \includegraphics[width=0.95\linewidth]{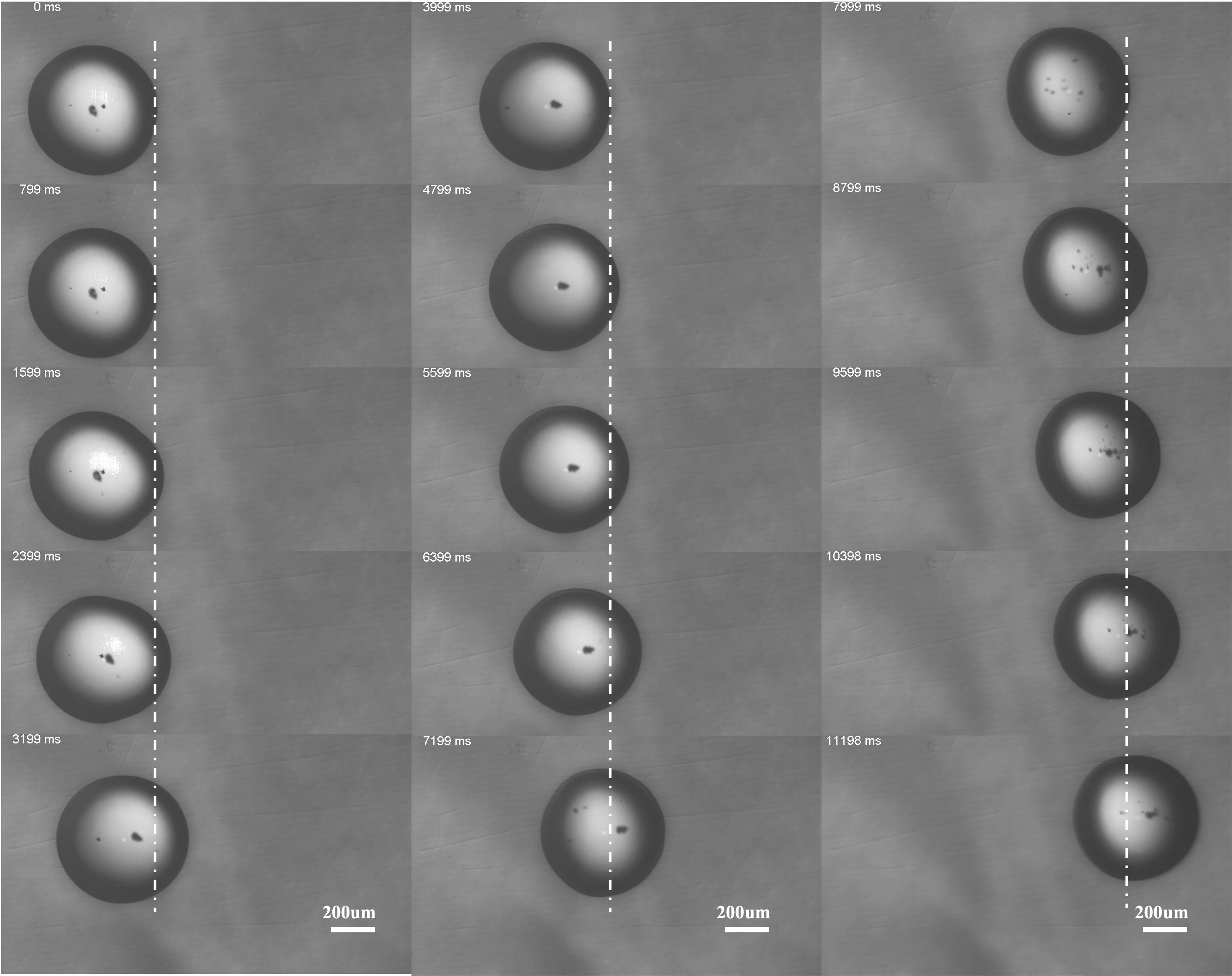}
  \caption{Top-view high-speed visualization of spontaneous droplet motion on a PMP substrate during evaporation.
  The image sequence was recorded using a Phantom high-speed camera and shows the lateral displacement of a sessile droplet at late stages of evaporation.
  To visualize internal motion during droplet translation, tracer particles used for particle image velocimetry (PIV) were dispersed inside the droplet prior to deposition.
  The dashed vertical line serves as a fixed spatial reference, highlighting the net lateral displacement of the droplet over time.
  The scale bar corresponds to $200~\mu$m.
  The full time-resolved dynamics are provided in Supplementary Video~1.
  }
  \label{fig:pmp_topview_motion}
\end{figure}

This interpretation is further supported by direct visualization of droplet motion on PMP substrates using high-speed imaging. Figure~\ref{fig:pmp_topview_motion} shows a sequence of top-view images captured by a Phantom high-speed camera, illustrating the lateral displacement of a droplet during the late stage of evaporation. To visualize internal flow and deformation during motion, tracer particles commonly used for particle image velocimetry (PIV) were dispersed inside the droplet prior to deposition. As evaporation proceeds, the droplet exhibits clear lateral translation accompanied by internal rearrangement, consistent with repeated depinning and re-pinning events of the three-phase contact line. The full time-resolved dynamics are provided in Supplementary Video~1 and 7.

\subsection{Kinematic scaling of the discharge event on PMP}

The results above establish that spontaneous motion emerges only after the droplet enters the late-stage charge-dissipative regime on PMP. To quantify the intensity of this motion, we analyzed the displacement and velocity of individual translation events and compared them with the relaxed charge fraction $1-Q_t/Q_i$. This analysis is motivated by the same physical question addressed in lubricated-surface Coulomb fission: once evaporation has driven the droplet into an electrostatically unstable regime, which observable best measures the strength of the relaxation event \cite{lin2026spontaneous}?

\begin{figure*}[!t]
  \centering
  \includegraphics[width=0.9\linewidth]{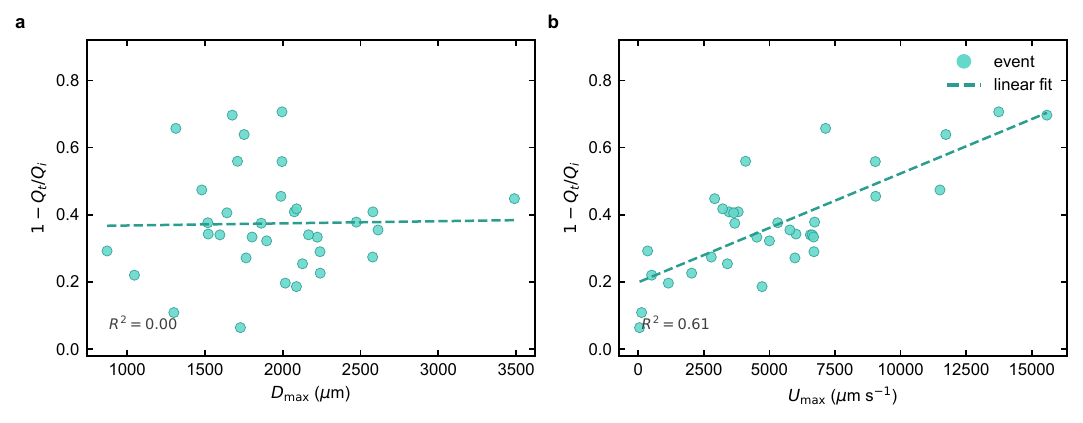}
  \caption{Kinematic signatures of charge relaxation during spontaneous translation on PMP, using code-tracked top-view events.
  (a) Relaxed charge fraction $1-Q_t/Q_i$ plotted against the maximum lateral displacement $D_{\max}$ of a translation event.
  (b) Relaxed charge fraction plotted against the maximum translational speed $U_{\max}$.
  Dashed lines are linear fits to the tracked event-level data.
  The displacement relation is weak because the final travel distance is strongly affected by contact-line re-pinning on the dry substrate, whereas the peak speed provides a cleaner measure of the rapid depinning/discharge segment.
  }
  \label{fig:kinematics_pmp}
\end{figure*}

We first consider the maximum lateral distance traveled during a discharge event. Figure~\ref{fig:kinematics_pmp}(a) shows essentially no collapse between $D_{\max}$ and $1-Q_t/Q_i$ for the code-tracked events ($R^2\simeq0.00$). This lack of collapse is physically expected for a dry solid substrate. Although a larger retained charge can provide a larger electrostatic driving force, the final displacement is an integrated outcome of many local interactions: nanoscale roughness, chemical inhomogeneity, and residual surface charge can all re-pin the contact line after motion begins. Consequently, $D_{\max}$ is sensitive not only to the electrostatic state of the droplet but also to the particular path encountered during a given event.

The maximum translational velocity $U_{\max}$ offers a more direct kinematic indicator of the discharge strength. Unlike the total distance, the peak speed is reached during the most mobile segment of the depinning event and is therefore less affected by later re-pinning. Figure~\ref{fig:kinematics_pmp}(b) shows that the relaxed charge fraction increases with $U_{\max}$ ($R^2\simeq0.61$ for the code-tracked events): droplets that move faster during the instability lose a larger fraction of their charge. This trend supports the interpretation that rapid whole-droplet translation is not merely a passive consequence of evaporation, but an active pathway for releasing electrostatic stress.

This distinction is mechanistically important. In the lubricated case, charge loss is accommodated through elongation, jetting, and Coulomb fission. Here, by contrast, the droplet remains intact and dissipates charge while executing lateral motion on a dry substrate. The kinematic data therefore support a two-step description of the PMP system: evaporation first increases the electrostatic fissility until a geometry-selected instability is reached, and the strength of the subsequent charge relaxation is encoded primarily in the peak speed of the translation event rather than in the final distance traveled.

\subsection{Comparison of long-term charge evolution on different substrates}

\begin{figure*}[!t]
  \centering
  \includegraphics[width=0.9\linewidth]{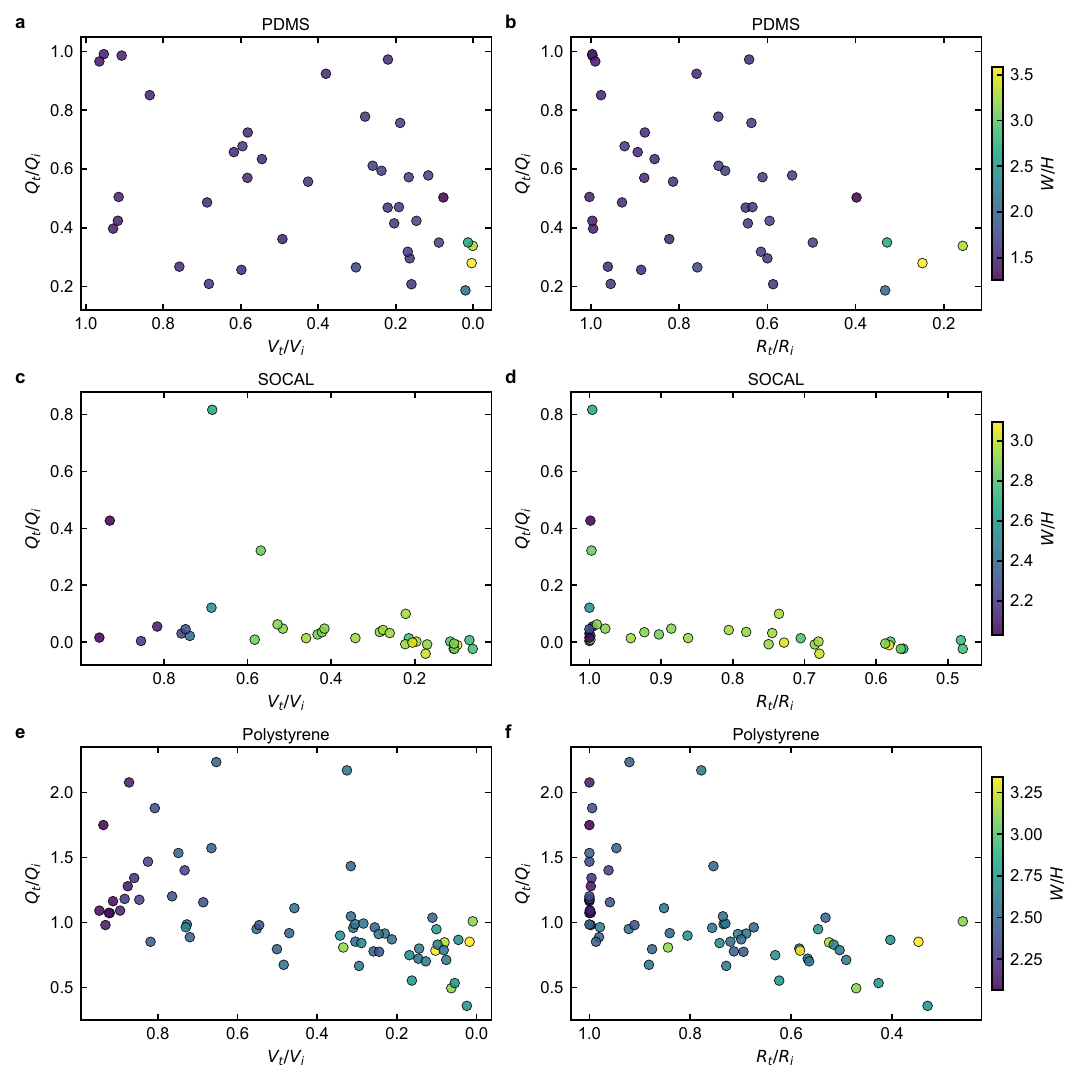}
  \caption{Comparison of charge evolution and droplet geometry during evaporation on different substrates.
  (a,c,e) Normalized droplet charge $Q_t/Q_i$ plotted as a function of the normalized droplet volume $V_t/V_i$ for PDMS, SOCAL, and polystyrene substrates, respectively.
  (b,d,f) Corresponding data plotted as a function of the normalized droplet radius $R_t/R_i$.
  In all panels, droplets have an initial volume of $0.6~\mu$L, and the color indicates the instantaneous droplet shape quantified by the width-to-height ratio $W/H$.
  While all substrates exhibit geometric compression at late stages of evaporation, only PMP (Fig.~\ref{fig:charge_evap_pmp}) shows a strong coupling between geometry, charge dissipation, and spontaneous droplet motion.
  }
  \label{fig:charge_evap_compare}
\end{figure*}

To place the PMP results in a broader context, the long-term evolution of droplet charge and geometry was examined on PDMS, SOCAL-coated surfaces, and polystyrene substrates under identical experimental conditions. Figure~\ref{fig:charge_evap_compare} summarizes the normalized charge $Q_t/Q_i$ as a function of both the normalized droplet volume $V_t/V_i$ and radius $R_t/R_i$ for all three substrates, with the color indicating the instantaneous droplet shape quantified by the width-to-height ratio $W/H$.

In contrast to PMP, none of the droplets on PDMS, SOCAL, or polystyrene exhibit spontaneous lateral motion throughout the evaporation process. However, the evolution of droplet geometry and charge differs markedly among these substrates. On PDMS (Fig.~\ref{fig:charge_evap_compare}(a,b)), the droplet remains stationary while the measured charge displays substantial scatter at all stages of evaporation. Pronounced droplet compression, reflected by an increase in $W/H$, emerges only at late stages ($V_t/V_i \approx 0.1$). Importantly, this geometric transition is not accompanied by a systematic change in $Q_t/Q_i$, indicating that droplet shape evolution and charge dynamics are weakly coupled on PDMS.

On SOCAL-coated substrates (Fig.~\ref{fig:charge_evap_compare}(c,d)), the droplet charge rapidly collapses to values near zero shortly after deposition and remains negligible throughout evaporation. This behavior confirms that SOCAL acts as an efficient charge sink. Although droplet compression is again observed at late stages ($V_t/V_i \approx 0.1$), the absence of measurable charge demonstrates that geometric evolution alone is insufficient to sustain charge retention or induce charge relaxation dynamics.

Polystyrene substrates exhibit yet another distinct behavior (Fig.~\ref{fig:charge_evap_compare}(e,f)). In this case, droplet compression occurs at relatively early stages of evaporation ($V_t/V_i \approx 0.5$). Simultaneously, the measured charge displays large variability and, in some instances, exceeds the initial charge ($Q_t/Q_i > 1$), indicating net charge transfer from the substrate to the droplet. This observation establishes polystyrene as an electrostatically active surface capable of injecting charge into the liquid during evaporation.

Taken together, these results demonstrate that while geometric compression during evaporation is a common feature across different substrates, only PMP supports a strong coupling between droplet geometry, charge retention, and spontaneous motion. The absence of droplet motion on PDMS, SOCAL, and polystyrene underscores that evaporation-induced shape changes alone are insufficient to trigger charge relaxation or mobility. Instead, the unique electrostatic passivity of PMP enables the emergence of a geometry-dependent instability that links evaporation, charge dissipation, and droplet motion. In this framework, substrate electrostatic character first determines whether a meaningful charge-retention pathway exists at all, droplet geometry then selects the onset of the instability, and the subsequent translational kinematics encode how strongly the droplet discharges once motion begins.

\section{Conclusions}

We have shown that the charge evolution of evaporating sessile droplets on dry polymer substrates is controlled jointly by substrate electrostatic character and droplet geometry. A zero-time Faraday-cup protocol establishes that PMP is the only tested substrate that preserves a reproducible initial charge, whereas PDMS produces heterogeneous charge exchange, SOCAL rapidly dissipates droplet charge, and polystyrene can inject charge into the liquid. This distinction is essential: without an electrostatically passive substrate, the initial state is not well defined and no geometry-controlled charge-retention pathway emerges.

On PMP, the droplet charge remains nearly constant during early evaporation and then collapses only after the droplet enters a small-volume state. Replotting the data in terms of $V_t/V_i$ reveals this transition more clearly than using the base radius, consistent with the idea that evaporation increases the droplet fissility as the size decreases. The measured contact-angle traces close the pinning part of the argument: spontaneous translation occurs only after the apparent angle approaches the pre-motion range and the electro-pinning number $\Pi_{\mathrm{dep}}$ exceeds the measured retention scale. Thus the onset is selected by the joint condition of high electro-capillary stress and depinning, not by geometry alone.

The post-threshold response differs fundamentally from the lubricated-surface Coulomb-fission pathway. Instead of elongating and emitting progeny droplets, charged droplets on dry PMP remain intact and release electrostatic stress through abrupt lateral translation. The total displacement is strongly affected by dry-surface pinning, whereas the peak translational velocity correlates more directly with the amount of charge relaxed during the event. These results identify spontaneous translation as a dry-substrate mode of evaporation-driven electrostatic discharge and establish substrate electrostatic passivity as a design parameter for controlling charged-droplet dynamics.

\section*{Supplementary Material}

Supplementary videos provide side-view and top-view recordings of the late-stage motion of charged droplets on PMP substrates, including repeated contact-line depinning and re-pinning events associated with spontaneous translation.

% \begin{acknowledgments}
% We wish to acknowledge the support of the Prof. Daniel Daniel.

% \end{acknowledgments}
\section*{AUTHOR DECLARATIONS}
Conflict of Interest: The authors have no conflicts to disclose.

Author Contributions: 
R. Xu: Data curation (equal); Methodology (equal); Formal analysis (equal); Investigation (equal);  Software (equal); Writing--original draft (equal); Writing--review \& editing (equal). Yanbo. Li and J. Zhang: Methodology (equal); Software (equal)
J. Wang and Yikai. Li: Methodology (equal); Conceptualization (equal); Formal analysis (equal);  Investigation (equal); Writing--original draft (equal); Writing--review \& editing (equal).

\section*{Data Availability Statement}
The data that support the findings of this study are available from the corresponding author upon reasonable request.

\section*{REFERENCES}
\bibliography{aipsamp}

@article{picknett1977evaporation,
  title = {The evaporation of sessile or pendant drops in still air},
  author = {Picknett, R. G. and Bexon, R.},
  journal = {Journal of Colloid and Interface Science},
  volume = {61},
  number = {2},
  pages = {336--350},
  year = {1977}
}

@article{deegan1997capillary,
  title = {Capillary flow as the cause of ring stains from dried liquid drops},
  author = {Deegan, Robert D. and Bakajin, Olgica and Dupont, Todd F. and Huber, Greb and Nagel, Sidney R. and Witten, Thomas A.},
  journal = {Nature},
  volume = {389},
  pages = {827--829},
  year = {1997},
  doi = {10.1038/39827}
}

@article{deegan2000contact,
  title = {Contact line deposits in an evaporating drop},
  author = {Deegan, Robert D. and Bakajin, Olgica and Dupont, Todd F. and Huber, Greb and Nagel, Sidney R. and Witten, Thomas A.},
  journal = {Physical Review E},
  volume = {62},
  number = {1},
  pages = {756--765},
  year = {2000}
}

@article{hu2002evaporation,
  title = {Evaporation of a sessile droplet on a substrate},
  author = {Hu, H. and Larson, R. G.},
  journal = {Journal of Physical Chemistry B},
  volume = {106},
  number = {6},
  pages = {1334--1344},
  year = {2002},
  doi = {10.1021/jp0118322}
}

@article{popov2005evaporative,
  title = {Evaporative deposition patterns: Spatial dimensions of the deposit},
  author = {Popov, Yuri O.},
  journal = {Physical Review E},
  volume = {71},
  pages = {036313},
  year = {2005},
  doi = {10.1103/PhysRevE.71.036313}
}

@article{stauber2015lifetimes,
  title = {On the lifetimes of evaporating droplets with related initial and receding contact angles},
  author = {Stauber, J. M. and Wilson, S. K. and Duffy, B. R. and Sefiane, K.},
  journal = {Physics of Fluids},
  volume = {27},
  pages = {122101},
  year = {2015},
  doi = {10.1063/1.4935232}
}

@article{wilson2023evaporation,
  title = {Evaporation of sessile droplets},
  author = {Wilson, Stephen K. and D'Ambrosio, Hannah-May},
  journal = {Annual Review of Fluid Mechanics},
  volume = {55},
  pages = {481--509},
  year = {2023},
  doi = {10.1146/annurev-fluid-031822-013213}
}

@article{baytekin2011mosaic,
  title = {The mosaic of surface charge in contact electrification},
  author = {Baytekin, H. Tarik and Patashinski, Alexander Z. and Branicki, Michal and Baytekin, Bilge and Soh, Siowling and Grzybowski, Bartosz A.},
  journal = {Science},
  volume = {333},
  number = {6040},
  pages = {308--312},
  year = {2011}
}

@article{choi2013spontaneous,
  title = {Spontaneous electrical charging of droplets by conventional pipetting},
  author = {Choi, Dongwhi and Lee, Hyunglae and Im, Dae Sung and Kang, In Seok and Lim, Geunbae and Kim, Dong Sung and Kang, Kwan Hyoung},
  journal = {Scientific Reports},
  volume = {3},
  pages = {2037},
  year = {2013}
}

@article{nauruzbayeva2020electrification,
  title = {Electrification at water--hydrophobe interfaces},
  author = {Nauruzbayeva, Jamilya and Sun, Zhongxu and Gallo, Anthony and Ibrahim, Mahmoud and Santamarina, J. Carlos and Mishra, Himanshu},
  journal = {Nature Communications},
  volume = {11},
  pages = {5285},
  year = {2020}
}

@article{li2022charge,
  title = {Spontaneous charging affects the motion of sliding drops},
  author = {Li, Xiaomei and Bista, Pravash and Stetten, Amy Z. and Bonart, Henning
            and Sch{\"u}r, Maximilian T. and Hardt, Steffen and Bodziony, Francisco
            and Marschall, Holger and Saal, Alexander and Deng, Xu and Berger, R{\"u}diger
            and Weber, Stefan A. L. and Butt, Hans-J{\"u}rgen},
  journal = {Nature Physics},
  volume = {18},
  pages = {713--719},
  year = {2022},
  doi = {10.1038/s41567-022-01563-6}
}

@article{lacks2011contact,
  title = {Contact electrification of insulating materials},
  author = {Lacks, Daniel J. and Sankaran, R. Mohan},
  journal = {Journal of Physics D: Applied Physics},
  volume = {44},
  pages = {453001},
  year = {2011},
  doi = {10.1088/0022-3727/44/45/453001}
}

@article{lin2022contactreview,
  title = {Contact electrification at the liquid--solid interface},
  author = {Lin, Shaoxin and Chen, Xiangyu and Wang, Zhong Lin},
  journal = {Chemical Reviews},
  volume = {122},
  pages = {5209--5232},
  year = {2022},
  doi = {10.1021/acs.chemrev.1c00176}
}

@article{diaz2023self,
  title = {Self-generated electrostatic forces of drops rebounding from hydrophobic surfaces},
  author = {D{\'\i}az, Diego and Li, Xiaomei and Bista, Pravash and Zhou, Xiaoteng
            and Darvish, Fahimeh and Butt, Hans-J{\"u}rgen and Kappl, Michael},
  journal = {Physics of Fluids},
  volume = {35},
  pages = {017111},
  year = {2023},
  doi = {10.1063/5.0130343}
}

@article{rayleigh1882liquid,
  title = {On the equilibrium of liquid conducting masses charged with electricity},
  author = {Rayleigh, Lord},
  journal = {Philosophical Magazine},
  volume = {14},
  pages = {184--186},
  year = {1882}
}

@article{gomez1994charge,
  title = {Charge and fission of droplets in electrostatic sprays},
  author = {Gomez, Alessandro and Tang, Keqi},
  journal = {Physics of Fluids},
  volume = {6},
  pages = {404--414},
  year = {1994}
}

@article{duft2003coulomb,
  title = {Rayleigh jets from levitated microdroplets},
  author = {Duft, D. and Achtzehn, T. and M{\"u}ller, R. and Huber, B. A. and Leisner, T.},
  journal = {Nature},
  volume = {421},
  pages = {128},
  year = {2003}
}

@article{lin2026spontaneous,
  title = {Spontaneous {Coulomb} fissions of drops on lubricated surfaces},
  author = {Lin, Marcus and Zhang, Peng and Ratschow, Aaron D. and Li, Oscar and Arunachalam, Sankara and Daniel, Dan},
  journal = {Proceedings of the National Academy of Sciences},
  volume = {123},
  number = {18},
  pages = {e2538161123},
  year = {2026},
  doi = {10.1073/pnas.2538161123}
}

@article{furmidge1962studies,
  title = {Studies at phase interfaces. {I}. The sliding of liquid drops on solid surfaces and a theory for spray retention},
  author = {Furmidge, C. G. L.},
  journal = {Journal of Colloid Science},
  volume = {17},
  number = {4},
  pages = {309--324},
  year = {1962}
}

@article{deGennes1985wetting,
  title = {Wetting: statics and dynamics},
  author = {de Gennes, P. G.},
  journal = {Reviews of Modern Physics},
  volume = {57},
  number = {3},
  pages = {827--863},
  year = {1985}
}

@article{joanny1984contact,
  title = {A model for contact angle hysteresis},
  author = {Joanny, J. F. and de Gennes, P. G.},
  journal = {Journal of Chemical Physics},
  volume = {81},
  pages = {552--562},
  year = {1984},
  doi = {10.1063/1.447337}
}

@article{eral2013contact,
  title = {Contact angle hysteresis: a review of fundamentals and applications},
  author = {Eral, H. B. and {'t~Mannetje}, D. J. C. M. and Oh, J. M.},
  journal = {Colloid and Polymer Science},
  volume = {291},
  pages = {247--260},
  year = {2013},
  doi = {10.1007/s00396-012-2796-6}
}

@article{takahashi1973measurement,
  title = {Measurement of electric charge of cloud droplets, drizzle, and raindrops},
  author = {Takahashi, Tsutomu},
  journal = {Reviews of Geophysics and Space Physics},
  volume = {11},
  number = {4},
  pages = {903--924},
  year = {1973},
  doi = {10.1029/RG011i004p00903}
}

@article{tinsley2000effects,
  title = {Effects of image charges on the scavenging of aerosol particles by cloud droplets and on droplet charging and possible ice nucleation processes},
  author = {Tinsley, B. A. and Rohrbaugh, R. P. and Hei, M. and Beard, K. V.},
  journal = {Journal of the Atmospheric Sciences},
  volume = {57},
  number = {13},
  pages = {2118--2134},
  year = {2000},
  doi = {10.1175/1520-0469(2000)057<2118:EOICOT>2.0.CO;2}
}

@article{tan2016enhanced,
  title = {Enhanced growth of single droplet by control of equivalent charge on droplet},
  author = {Tan, Xiao and Qiu, Yunhao and Yang, Yong and Liu, Dawei and Lu, XinPei and Pan, Yuan},
  journal = {IEEE Transactions on Plasma Science},
  volume = {44},
  number = {11},
  pages = {2724--2728},
  year = {2016},
  doi = {10.1109/TPS.2016.2608832}
}

@article{iribarne1976evaporation,
  title = {On the evaporation of small ions from charged droplets},
  author = {Iribarne, J. V. and Thomson, B. A.},
  journal = {Journal of Chemical Physics},
  volume = {64},
  number = {6},
  pages = {2287--2294},
  year = {1976},
  doi = {10.1063/1.432536}
}

@article{fenn1993ion,
  title = {Ion formation from charged droplets: Roles of geometry, energy, and time},
  author = {Fenn, John B.},
  journal = {Journal of the American Society for Mass Spectrometry},
  volume = {4},
  number = {7},
  pages = {524--535},
  year = {1993},
  doi = {10.1016/1044-0305(93)85014-O}
}

@article{kebarle2000mechanisms,
  title = {On the mechanisms by which the charged droplets produced by electrospray lead to gas phase ions},
  author = {Kebarle, Paul and Peschke, Michael},
  journal = {Analytica Chimica Acta},
  volume = {406},
  number = {1},
  pages = {11--35},
  year = {2000},
  doi = {10.1016/S0003-2670(99)00598-X}
}

@article{sun2016solid,
  title = {Solid-to-liquid charge transfer for generating droplets with tunable charge},
  author = {Sun, Yajuan and Huang, Xu and Soh, Siowling},
  journal = {Angewandte Chemie International Edition},
  volume = {55},
  number = {34},
  pages = {9956--9960},
  year = {2016},
  doi = {10.1002/anie.201604378}
}

@article{miljkovic2013electrostatic,
  title = {Electrostatic charging of jumping droplets},
  author = {Miljkovic, Nenad and Preston, Daniel J. and Enright, Ryan and Wang, Evelyn N.},
  journal = {Nature Communications},
  volume = {4},
  pages = {2517},
  year = {2013},
  doi = {10.1038/ncomms3517}
}

@article{sun2019surface,
  title = {Surface charge printing for programmed droplet transport},
  author = {Sun, Qiangqiang and Wang, Dehui and Li, Yanan and Zhang, Jiahui and Ye, Shuji and Cui, Jiaxi and Chen, Longquan and Wang, Zuankai and Butt, Hans-J{\"u}rgen and Vollmer, Doris and Deng, Xu},
  journal = {Nature Materials},
  volume = {18},
  pages = {936--941},
  year = {2019},
  doi = {10.1038/s41563-019-0440-2}
}

@article{xu2021impact,
  title = {Impact dynamics of a charged droplet onto different substrates},
  author = {Xu, Haojie and Wang, Junfeng and Wang, Zhentao and Yu, Kai and Xu, Huibin and Wang, Dongbao and Zhang, Wei},
  journal = {Physics of Fluids},
  volume = {33},
  pages = {102111},
  year = {2021},
  doi = {10.1063/5.0066381}
}

@article{jung2008electrical,
  title = {Electrical charging of a conducting water droplet in a dielectric fluid on the electrode surface},
  author = {Jung, Yong-Mi and Oh, Hyun-Chang and Kang, In Seok},
  journal = {Journal of Colloid and Interface Science},
  volume = {322},
  number = {2},
  pages = {617--623},
  year = {2008},
  doi = {10.1016/j.jcis.2008.04.019}
}

@article{im2011electrophoresis,
  title = {Electrophoresis of a charged droplet in a dielectric liquid for droplet actuation},
  author = {Im, Do Jin and Noh, Jihoon and Moon, Dustin and Kang, In Seok},
  journal = {Analytical Chemistry},
  volume = {83},
  number = {13},
  pages = {5168--5174},
  year = {2011},
  doi = {10.1021/ac200248x}
}

@article{im2012discrete,
  title = {Discrete electrostatic charge transfer by the electrophoresis of a charged droplet in a dielectric liquid},
  author = {Im, Do Jin and Ahn, Myung Mo and Yoo, Byeong Sun and Moon, Dustin and Lee, Dong Woog and Kang, In Seok},
  journal = {Langmuir},
  volume = {28},
  number = {31},
  pages = {11656--11661},
  year = {2012},
  doi = {10.1021/la3014392}
}

@article{li2005charge,
  title = {Charge limits on droplets during evaporation},
  author = {Li, Kuo-Yen and Tu, Haohua and Ray, Asit K.},
  journal = {Langmuir},
  volume = {21},
  number = {9},
  pages = {3786--3794},
  year = {2005},
  doi = {10.1021/la047973n}
}

@article{zilch2008charge,
  title = {Charge separation in the aerodynamic breakup of micrometer-sized water droplets},
  author = {Zilch, Lloyd W. and Maze, Joshua T. and Smith, John W. and Ewing, George E. and Jarrold, Martin F.},
  journal = {Journal of Physical Chemistry A},
  volume = {112},
  number = {51},
  pages = {13352--13363},
  year = {2008},
  doi = {10.1021/jp806995h}
}
\end{document}